Commentary: analyzing binary data using MCPMod when zero counts are expected

Yi Liu, Sebastian Bossert, Rui Wu, Dooti Roy, Frank Fleischer, Qiqi Deng

1. **Background and methodology**

Bretz et al (2005) [1] proposed multiple Comparison Procedure and Modeling (MCPMod) method to design and analyze dose-finding study. Pinheiro (2014)[2] then generalized it to various types of endpoint, including but not limited to binary endpoint, survival endpoint, count data, and longitudinal data. The generalization has significantly expanded the applicability of MCPMod, and has been widely adopted in practice.

For binary data, Pinheiro (2013)[2] proposed to use logistic regression, and build the generalized MCPMod on the logit scale:
$$Y_{ij} = \text{logit}(X_{ij}) = \log\left(\frac{X_{ij}}{n_i - X_{ij}}\right) \sim N\left(\log\left(\frac{p_i}{1-p_i}\right), \frac{1}{n_i p_i(1-p_i)}\right), \text{ for } i=1, \ldots, D, \text{ and } j=1,\ldots, n_i, \quad (1),$$
where $X_{ij}$ is number of responder for subject $j$ taking dose $i$.

If $M$ models are used in candidate set, the optimal contrast is inversely proportional to the standard error:
$$\boldsymbol{C}_m^{opt} \propto \boldsymbol{S}_m^{-1}\left(\boldsymbol{\mu}_m^0 - \frac{\boldsymbol{\mu}_m^{0\prime} \boldsymbol{S}_m^{-1} \boldsymbol{1}}{\boldsymbol{1}' \boldsymbol{S}_m^{-1} \boldsymbol{1}}\right), \text{ for } m=1,\ldots M, \quad (2)$$
where $\boldsymbol{\mu}_m^0 = \log\left(\frac{\boldsymbol{p}_m}{1-\boldsymbol{p}_m}\right)$, and $\boldsymbol{p}_m$ indicates the vector of $p_i$ under model $m$.

Pinheiro (2013)[2] recommended to use the estimated covariance matrix from the observed data to recalculate the optimal contrast and the critical value of the test:
$$\boldsymbol{S}_m^{-1} = \hat{\boldsymbol{S}}^{-1} \quad (3)$$
This is also the default analysis approach when using the DoseFinding package [3] to analyze the data.

However, when this approach is used in practice, we discovered that it can lead to problematic results for cases when there are zero counts in at least one arm. There are two issues. First, the estimation of regression coefficients from logistic regression is unstable. In addition, the updated contrast coefficients for MCPMod testing can be inefficient, and sometimes not sensible.

For many phase II studies it is common to have small sample sizes per arm with low placebo response rates jointly. Under such circumstances, it cannot be excluded to have a zero count observed. For example, when the placebo response rate is 10%, there is about 4% chance to observe zero responders in the placebo group (Table 1), or other dose group(s), which has a similar response rate as placebo. This probability increases to 35% if you have an arm of only 10 subjects planned. When the placebo response rate is 5%, the numbers could be as high as 21% for 30 patients, and 60% for 10 patients.

Table 1: Probability of zero count in placebo arm under different sample sizes, assuming 10% or 5% response rate

| Sample size on placebo arm = | 10 | 15 | 20 | 30 |
|---|---|---|---|---|
| Probability of zero count (p=10%) | 0.35 | 0.21 | 0.12 | 0.04 |
| Probability of zero count (p=5%) | 0.60 | 0.46 | 0.36 | 0.21 |

In this manuscript, we would like to illustrate the potential problem of (3) using a case study and simulations. An alternative method using logistic regression was evaluated to get a stable estimate of response for each dose group. In addition, we evaluated two options to address the second issue with problematic contrast coefficients:

A. Calculate the contrast coefficients using the covariance matrix from expected response rate of the pre-specified candidate set models. For binary data, since the variance is a function of the response rate as shown in (1), the different dose response candidate set models not only have different $\boldsymbol{\mu}_m^0$, but also have different variance matrices
$$S_m^{-1} = \mathrm{diag}(N)\boldsymbol{p}_m(1-\boldsymbol{p}_m), \quad (4)$$
Where $N$ is the vector of $n_i$. The optimal contrast can be the calculated using (2) and (4).

B. Calculate the contrast coefficients using a weight proportional to sample size ($N$) or equivalently allocation ratio:
$$S_m^{-1} \propto \mathrm{diag}(N), \quad (5)$$

## 2. A case study

Suppose there is phase II study to evaluate a candidate compound to treat patients with plaque psoriasis. The primary endpoint is binary: whether the patient has at least 75% reduction from baseline for PASI score (PASI 75) [4] at week 12. There are five dose groups tested, placebo, 0.125mg, 0.25mg, 0.5mg and 1mg. The maximum treatment effect is expected to be achieved at the highest dose with 40% increase in response rate over placebo. With 30 patients per arm, and expected placebo rate of 5-10%, the study will have about >90% of power (under one-sided α=5%) to detect a non-flat dose response relationship if a candidate set models as in Figure 1 are used for MCPMod.

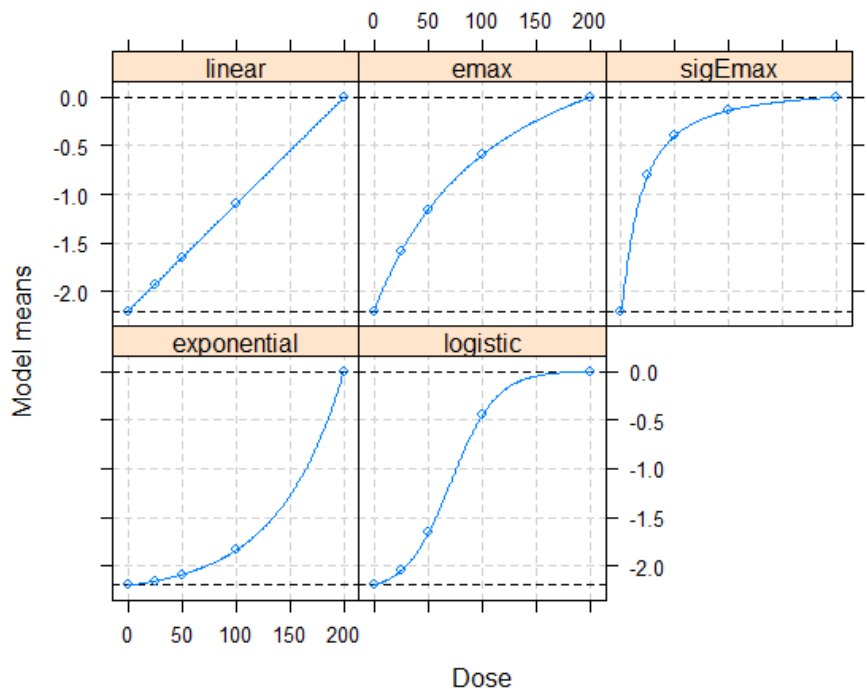

Figure 1: Candidate set models for MCPMod analysis in logit scale.

When there is a zero count occurring, the results from MCPMod based on logistic regression can be problematic. For example suppose the observed number of responders for the 5 groups are 0, 13, 14, 15, 15 respectively. It is obvious that such a result should lead to a clear rejection of the null hypothesis. This is reflected on the extremely small p-value from chi-square trend test in proportions [5, 6] (p-value=0.00163) or conditional exact Cochran-Armitage test [7, 8] (p-value=0.000955). The fisher's exact test comparing the lowest dose against placebo can also easily detect it (p-value = 0.000159).

However, logistic regression will be problematic in this case as shown in Table 2. The first apparent issue is that as the log odds of zero is negative infinity, the estimated log odds or odds ratio in logistic regression becomes very unstable which means that a huge standard error is induced. This is also the case when the observed response rate equals one. For simplicity, we will only focus on zero count scenario here.

There has been various methods developed to handle this rare events situation in regression [9-13]. Among those that have been implemented in commonly used tools (R, SAS® software), Firth's penalized logistic regression [13] is one of the methods that perform better than the others [14]. In this manuscript, Firth's penalized regression is used in subsequent investigations to handle the zero count observations in a dose finding study setting. In Table 3, it is clear that there is a significant treatment effect for each active dose against placebo.

However, even if the coefficient from Firth's penalized logistic regression is used in estimating the response ($\text{logit}(p_i)$) from each dose group and the associated standard error, the smallest p-value from MCPMod is still 0.071 and it fails to reject a non-flat dose response relationship as shown in Table 4.

This puzzling and counter-intuitive result is due to calculation of the updated optimal contrast. Table 4 showed that the contrast coefficients of the placebo arm are close to zero for all models except sigEmax. When the contrast coefficient goes to zero, the data from the placebo arm is actually not contributing to the test. In this case, the contrast test is essentially testing whether there is any difference among the four active dose groups, instead of whether some of the active doses are different from placebo. Additionally, even though the contrast value under sigEmax is non-zero for placebo, it is still small compared to the other arms. Therefore there is a heavy down-weighting of the placebo results which in turn leads to a non-significant p-value for the MCPMod test both overall and also for the specific models.

The issue described above about the contrast coefficients occurs, when the estimated covariance matrix from observed data is used to recalculate optimal contrast. As shown in (2), optimal contrast is inversely proportional to standard error. When the estimated covariance matrix $\hat{S}$ as shown in Table 5 is used to update the optimal contrast following (3), the big standard error in the placebo arm consequently leads to a small contrast value on placebo arm, which down weights or even essentially excludes the placebo data from the analysis.

As a comparison, the modified MCPMod analysis results based on A and B in section 1 are applied to the same data displayed in Table 6.

From the above results, we can see that p-value for approach B is smaller than that of approach A and the contrast coefficient for the placebo arm is also larger in approach B. One explanation is that approach A down-weights the data from arms with smaller response rate, and that may still lead to an inefficient contrast when the response rate is expected to be very low (<3%). In the later section, the simulation study will provide more details and clarity.

### 3. Simulation

A simulation study is conducted to evaluate the impact on type I error and power using the set up mentioned in section 1. For the analysis of simulated study data, when there is zero count observed in any arm, Firth's penalized logistic regression will be used. Otherwise, the ordinary logistic regression is used.

Five approaches displayed in Table 7 are compared in the following. In addition, the probability of having a zero count in at least one arm is provided. The results are displayed in Table 8-11.

Table 7: Summary of methods investigated in simulation study

| Label | Methods |
| --- | --- |
| "Orig" | MCPMod, with contrast calculated using $S_m^{-1} = \hat{S}^{-1}$ |
| A | MCPMod, with contrast calculated using $S_m^{-1} = \text{diag}(N)p_m(1 - p_m)$, |
| B | MCPMod, with contrast calculated using $S_m^{-1} \propto \text{diag}(N)$ |
| Trend | Chi-square trend test as implemented by prop.trend.test function in R |
| exact.CA | Exact Cochran-Armitage test as implemented by catt_exact() function from CATTexact package in R |

Table 8 evaluated the type I error under the null hypothesis of no treatment effect, with the true placebo response rate ranging from 1%-70%. A one-sided $\alpha = 0.05$ and the candidate models in Figure 1 are used for analysis. When the response rate is between 1% to 10%, the probabilities of observing zero in any arm (P(zero)) range from 20% to close to 100%. And when response rate is 30% or higher, zero counts rarely occur. All methods control type I error around or under the nominal level across all scenarios. Under low response rate (<10% or less), some of the methods are even conservative with actual type I error lower than nominal level. This is likely because normal approximation is not accurate when number of events is small. We also tested the case when candidate set model used the correct placebo rate in table 1, which yielded similar results.

Table 9 displayed the power when placebo response rate =10%. Under this response rate, there is about 4% chance of observing zero for placebo arm. Depending on the shape of the dose response curve, the overall probability of observing zero response in any arm could be as high as 8-12% under scenario #4-6 when some of the active doses are similar to placebo (e.g exponential shape). When the effect size is large and the model is correct (#1-5), the power is higher than 95% and all approaches are comparable. For all scenarios, the original MCPMod (which updates the contrast based on observed data) yields comparable power with trend test and exact Cochran-

Armitage test, but didn't have power advantage over A and B. Actually A and B leads to more consistent and higher power than the original method, when the effect size is moderate and the model is not accurate (#6-8).

The idea of updating optimal contrast based on observed data seems to be attractive at the first glance. Since $C^{opt} \propto S^{-1}(\mu_m^0 - \frac{\mu_m^{0'} S^{-1} 1}{1' S^{-1} 1})$, providing a more accurate S matrix theoretically could improve the power. But the caveat is that to avoid the type I error, $\hat{\mu}_m^0$ can not be used to update the contrast. With sparse data, the logistic-based regression model tends to give a relatively big estimate of the effect size ($\hat{\beta}$) with a big standard error. If $\hat{\mu}_m^0$ can also be used to update the contrast together with $\hat{S}$, the effect of increased S values will be compensated by increased values of $\mu_m^0$, and the placebo arm may still get a reasonable weight. However, since $\hat{\mu}_m^0$ cannot be used to update the contrast. The increase in standard error alone has such a dramatic effect that the weight of an arm with zero count diminishes.

Table 10 displayed the power when the placebo response rate equals 30%. Under this response rate, zero counts do rarely occur. MCPMod based approaches (original, A and B) provide similar power under all scenarios, and are more powerful than trend test or exact Cochran-Armitage test. Again, A and B provides similar or even slightly higher power than the original MCPMod approach.

Table 11 displayed the power when the placebo response rate equals 1%. This extreme set up is used to magnify the impact of the different contrasts on the results of MCPMod. In this case B provides comparable power to the trend test and the exact Cochran-Armitage test. But the original MCPMod and A have low power when the effect size is moderate and the model is not accurate (#6-8), especially for an emax-like shape of curve that all active doses are similar (#7-8). This is due to the problem of down weighting placebo data as illustrated in section 1.

In each of the set-ups and in order to evaluate the robustness of the results, the investigation is repeated when the placebo rate is mis-specified (labeled as "shifted model"), e.g. the placebo rate is assumed to be 10% at the design stage, but the true rate is actually 35% (see Table 9). Scenarios when true model is not included in candidate model is also considered We also look into the cases where the randomization ratio is not balanced. These investigations demonstrated similar behavior in the sense that method B performs similarly or better than the original approach.

In summary, the MCPMod method combines hypothesis testing and modeling of dose ranging trials under one umbrella, and is an efficient way to analyzing data from trials with multiple doses. However, when zero counts are expected for binary data, simply pre-specifying that optimal contrasts will be derived/updated based on the observed data may lead to problematic results. In this case, one work-around is to use a two-step mechanism: first apply penalized logistic regression or other modified logistic regression model to obtain a reasonable estimate of effect size and corresponding standard error. As a second step apply MCPMod with contrasts that are derived based on the response rate from different candidate set model, and simply adjusted by allocation ratio (method B). This approach produces reasonable and robust outcome for observed data cases with or without zero counts across all investigated scenarios.

Table 2: Model fitting results of logistic regression

|  | Estimate | Std.Error | P-value |
|---|---|---|---|
| Intercept | -18.57 | 1190.87 | 0.988 |
| 0.125 | 18.02 | 1190.87 | 0.988 |
| 0.25 | 17.87 | 1190.87 | 0.988 |
| 0.5 | 18.16 | 1190.87 | 0.988 |
| 1 | 18.16 | 1190.87 | 0.988 |

Table 3: Model fitting results of firth penalized logistic regression

|  | Estimate | Std.Error | P-value |
|---|---|---|---|
| Intercept | -4.11 | 1.45 | $7.6*10^{-10}$ |
| 0.125 | 3.58 | 1.50 | $1.12*10^{-4}$ |
| 0.25 | 3.44 | 1.50 | $2.87*10^{-4}$ |
| 0.5 | 3.72 | 1.50 | $4.26*10^{-5}$ |
| 1 | 3.72 | 1.50 | $4.26*10^{-5}$ |

Table 4: MCPMod analysis results based on Firth's penalized logistic regression, with contrast coefficients weighted by inverse of observed covariance matrix

| Contrast coefficients: | | | | | |
|---|---|---|---|---|---|
|  | linear | Emax | sigEmax | exponential | logistic |
| 0 | -0.047 | -0.076 | -0.195 | -0.026 | -0.046 |
| 0.125 | -0.502 | -0.611 | -0.698 | -0.353 | -0.591 |
| 0.25 | -0.306 | -0.246 | -0.054 | -0.303 | -0.351 |
| 0.5 | 0.049 | 0.214 | 0.363 | -0.184 | 0.357 |
| 1 | 0.806 | 0.718 | 0.584 | 0.866 | 0.631 |

| Multiple contrast test: | | |
|---|---|---|
|  | t-stat | Adjust p-value |
| sigEmax | 1.799 | 0.071 |
| Emax | 1.112 | 0.228 |
| logistic | 0.917 | 0.295 |
| linear | 0.858 | 0.317 |
| exponential | 0.605 | 0.417 |

Table 5: Covariance matrix from Firth's penalized logistic regression

| Dose | 0 | 0.125 | 0.25 | 0.5 | 1 |
|---|---|---|---|---|---|
| 0 | 2.101 | 0 | 0 | 0 | 0 |
| 0.125 | 0 | 0.143 | 0 | 0 | 0 |
| 0.25 | 0 | 0 | 0.149 | 0 | 0 |
| 0.5 | 0 | 0 | 0 | 0.139 | 0 |
| 1 | 0 | 0 | 0 | 0 | 0.139 |

Table 6: MCPMod analysis results based on Firth's penalized logistic regression, a) with contrast coefficients weighted by inverse of covariance matrix in design stage; b) with contrast coefficients weighted by allocation ratio

a)

| Contrast coefficients: | | | | | |
|---|---|---|---|---|---|
| Model<br>Dose | linear | Emax | sigEmax | exponential | logistic |
| 0 | -0.248 | -0.321 | -0.530 | -0.219 | -0.317 |
| 0.125 | -0.265 | -0.360 | -0.454 | -0.221 | -0.337 |
| 0.25 | -0.259 | -0.267 | -0.006 | -0.225 | -0.361 |
| 0.5 | -0.116 | 0.124 | 0.385 | -0.229 | 0.242 |
| 1 | 0.888 | 0.825 | 0.604 | 0.894 | 0.773 |
| Multiple contrast test: | | | | | |

| | t-stat | Adjust p-value |
|---|---|---|
| sigEmax | 2.448 | 0.013 |
| Emax | 2.279 | 0.020 |
| logistic | 2.252 | 0.021 |
| linear | 2.019 | 0.037 |
| exponential | 1.876 | 0.051 |

b)

| Contrast coefficients: | | | | | |
|---|---|---|---|---|---|
| Model<br>Dose | linear | Emax | sigEmax | exponential | logistic |
| 0 | -0.474 | -0.640 | -0.839 | -0.290 | -0.470 |
| 0.125 | -0.316 | -0.280 | -0.057 | -0.266 | -0.392 |
| 0.25 | -0.158 | -0.028 | 0.175 | -0.231 | -0.198 |
| 0.5 | 0.158 | 0.301 | 0.322 | -0.098 | 0.415 |
| 1 | 0.791 | 0.648 | 0.399 | 0.885 | 0.644 |
| Multiple contrast test: | | | | | |

| | t-stat | Adjust p-value |
|---|---|---|
| emax | 2.499 | 0.011 |
| sigEmax | 2.498 | 0.011 |
| logistic | 2.451 | 0.013 |
| linear | 2.429 | 0.014 |
| exponential | 2.135 | 0.028 |

Table 8: Actual Type I error under flat dose response curve when nominal α=5%.

| | Simulation response rate | P(zero) | Orig. | A | B | Trend | exact. CA |
|---|---|---|---|---|---|---|---|
| Null model | 0.01, 0.01, 0.01, 0.01, 0.01* | 0.9992 | 0.0019 | 0.0019 | 0.0019 | 0.0355 | 0.0188 |
| | 0.1, 0.1, 0.1, 0.1, 0.1 | 0.1966 | 0.0234 | 0.0432 | 0.0359 | 0.0499 | 0.0404 |
| | 0.3, 0.3, 0.3, 0.3, 0.3 | 0.0030 | 0.0381 | 0.0495 | 0.0451 | 0.0478 | 0.0413 |
| | 0.5, 0.5, 0.5, 0.5, 0.5 | 0 | 0.0451 | 0.0494 | 0.0499 | 0.0576 | 0.0480 |
| | 0.7, 0.7, 0.7, 0.7, 0.7 | 0 | 0.0425 | 0.0456 | 0.0478 | 0.0478 | 0.0451 |

*Out of 10000 simulation runs, 2020 have zero responder in all dose groups. Type-I error summary is based on 7980 runs with at least one responder in any dose group.

Table 9: Power when placebo response rate =10%, and maximum treatment effect is up to 40%.

| | No. | Simulation response rate | P(zero) | Orig. | A) | B) | Trend | exact. CA |
|---|---|---|---|---|---|---|---|---|
| Correct model: Linear, emax, sigEmax, exp, logistic | 1 | 0.1, 0.13, 0.16, 0.25, 0.5 | 0.0641 | 0.9839 | 0.9901 | 0.989 | 0.9715 | 0.9841 |
| | 2 | 0.1, 0.17, 0.24, 0.36, 0.5 | 0.0483 | 0.9811 | 0.9867 | 0.9861 | 0.9796 | 0.9888 |
| | 3 | 0.1, 0.31, 0.4, 0.47, 0.5 | 0.0445 | 0.9663 | 0.9769 | 0.9837 | 0.956 | 0.9759 |
| | 4 | 0.1, 0.1, 0.11, 0.14, 0.5 | 0.1225 | 0.9905 | 0.9956 | 0.9946 | 0.9617 | 0.9767 |
| | 5 | 0.1, 0.11, 0.16, 0.39, 0.5 | 0.0783 | 0.9947 | 0.9968 | 0.9965 | 0.9939 | 0.9961 |
| Moderate "linear" | 6 | 0.1, 0.1, 0.15, 0.2, 0.25 | 0.0922 | 0.4786 | 0.5709 | 0.5476 | 0.4725 | 0.5701 |
| Moderate Plateau | 7 | 0.1, 0.3, 0.3, 0.3, 0.3 | 0.0447 | 0.3548 | 0.4501 | 0.5669 | 0.3392 | 0.4409 |
| Strong plateau | 8 | 0.1, 0.5, 0.5, 0.5, 0.5 | 0.0444 | 0.8887 | 0.9343 | 0.9873 | 0.8426 | 0.8992 |
| Shifted model: Linear, emax, sigEmax, exp, logistic | 9 | 0.35, 0.38, 0.41, 0.5, 0.75 | 0 | 0.9672 | 0.9734 | 0.9695 | 0.9148 | 0.9482 |
| | 10 | 0.35, 0.42, 0.49, 0.61, 0.75 | 0 | 0.9638 | 0.9658 | 0.9655 | 0.9481 | 0.9708 |
| | 11 | 0.35, 0.56, 0.65, 0.72, 0.75 | 0 | 0.9603 | 0.9483 | 0.9645 | 0.9376 | 0.9621 |
| | 12 | 0.35, 0.35, 0.36, 0.39, 0.75 | 0 | 0.9781 | 0.9843 | 0.981 | 0.8592 | 0.9103 |
| | 13 | 0.35, 0.36, 0.41, 0.64, 0.75 | 0 | 0.9865 | 0.9893 | 0.9878 | 0.9757 | 0.9864 |
| Misspec. model: betaMod, sigEmax, logistic | 14 | 0.1, 0.25, 0.37, 0.5, 0.26 | 0.0448 | 0.7947 | 0.7307 | 0.7968 | 0.589 | 0.6933 |
| | 15 | 0.1, 0.45, 0.48, 0.49, 0.5 | 0.0444 | 0.9172 | 0.9468 | 0.9859 | 0.8772 | 0.9269 |
| | 16 | 0.1, 0.11, 0.14, 0.42, 0.5 | 0.0819 | 0.9958 | 0.998 | 0.9979 | 0.995 | 0.9967 |
| Misspec shifted model: betaMod, sigEmax, logistic | 17 | 0.35, 0.5, 0.62, 0.75, 0.51 | 0 | 0.6786 | 0.6704 | 0.7447 | 0.5328 | 0.6260 |
| | 18 | 0.35, 0.7, 0.73, 0.74, 0.75 | 0 | 0.9659 | 0.8852 | 0.9675 | 0.8727 | 0.9200 |
| | 19 | 0.35, 0.36, 0.39, 0.67, 0.75 | 0 | 0.9917 | 0.9935 | 0.9927 | 0.981 | 0.9904 |

* 10000 simulation runs for all results, beside of exact CA for which only 1000 simulations were conducted due to extensive run time.

Table 10: Power when placebo response rate =30%, and maximum treatment effect is up to 40%.

| | No. | Simulation response rate | P(zero) | Orig. | A) | B) | Trend | exact. CA |
|---|---|---|---|---|---|---|---|---|
| Correct model: Linear, emax, sigEmax, exp, logistic | 1 | 0.3, 0.35, 0.4, 0.5, 0.7 | 0 | 0.9623 | 0.9650 | 0.9654 | 0.9303 | 0.9559 |
| | 2 | 0.3, 0.41, 0.49, 0.6, 0.7 | 0 | 0.9597 | 0.9617 | 0.9624 | 0.9493 | 0.9700 |
| | 3 | 0.3, 0.56, 0.63, 0.68, 0.7 | 0 | 0.9601 | 0.9619 | 0.9631 | 0.9182 | 0.9491 |
| | 4 | 0.3, 0.31, 0.32, 0.36, 0.7 | 0 | 0.9725 | 0.9758 | 0.9758 | 0.8718 | 0.9180 |
| | 5 | 0.3, 0.33, 0.39, 0.62, 0.7 | 0 | 0.9860 | 0.9874 | 0.9874 | 0.9784 | 0.9875 |
| Moderate "linear" | 6 | 0.3, 0.3, 0.35, 0.4, 0.45 | 0 | 0.3715 | 0.3973 | 0.3963 | 0.3032 | 0.3887 |
| Moderate Plateau | 7 | 0.3, 0.5, 0.5, 0.5, 0.5 | 0 | 0.4276 | 0.4518 | 0.4586 | 0.2982 | 0.3876 |
| Strong plateau | 8 | 0.3, 0.7, 0.7, 0.7, 0.7 | 0 | 0.9658 | 0.9655 | 0.9677 | 0.8321 | 0.8893 |
| Shifted model: Linear, emax, sigEmax, exp, logistic | 9 | 0.1, 0.15, 0.2, 0.3, 0.5 | 0.0529 | 0.9807 | 0.9858 | 0.9860 | 0.9747 | 0.9863 |
| | 10 | 0.1, 0.21, 0.29, 0.4, 0.5 | 0.0453 | 0.9777 | 0.9833 | 0.9841 | 0.9766 | 0.9872 |
| | 11 | 0.1, 0.36, 0.43, 0.48, 0.5 | 0.0444 | 0.9549 | 0.9843 | 0.9841 | 0.9359 | 0.9642 |
| | 12 | 0.1, 0.11, 0.12, 0.16, 0.5 | 0.0983 | 0.9874 | 0.9921 | 0.9922 | 0.9608 | 0.9778 |
| | 13 | 0.1, 0.13, 0.19, 0.42, 0.5 | 0.0614 | 0.9934 | 0.9956 | 0.9957 | 0.9927 | 0.9959 |
| Misspec. model: betaMod, sigEmax, logistic | 14 | 0.3, 0.5, 0.61, 0.7, 0.51 | 0 | 0.7842 | 0.8189 | 0.8147 | 0.7842 | 0.6953 |
| | 15 | 0.3, 0.67, 0.68, 0.69, 0.7 | 0 | 0.9607 | 0.9607 | 0.9634 | 0.9607 | 0.8997 |
| | 16 | 0.3, 0.31, 0.37, 0.64, 0.7 | 0.0001 | 0.9911 | 0.9927 | 0.9922 | 0.9911 | 0.9913 |
| Misspec. shifted model: betaMod, sigEmax, logistic | 17 | 0.1, 0.3, 0.41, 0.5, 0.31 | 0.0445 | 0.8184 | 0.8844 | 0.8806 | 0.6483 | 0.7479 |
| | 18 | 0.1, 0.47, 0.48, 0.49, 0.5 | 0.0444 | 0.9023 | 0.9846 | 0.9855 | 0.8605 | 0.9144 |
| | 19 | 0.1, 0.11, 0.17, 0.44, 0.5 | 0.0773 | 0.9959 | 0.9976 | 0.9976 | 0.9956 | 0.9974 |

Table 11: Power when placebo response rate =1%, and maximum treatment effect is up to 40%.

| | No. | Simulation response rate | P(zero) | Orig. | A) | B) | Trend | exact. CA |
|---|---|---|---|---|---|---|---|---|
| Correct model: Linear, emax, sigEmax, exp, logistic | 1 | 0.01, 0.017, 0.028, 0.077, 0.41 | 0.9465 | 0.9988 | 0.9995 | 0.9997 | 0.9993 | 0.9996 |
| | 2 | 0.01, 0.032, 0.07, 0.18, 0.41 | 0.8564 | 0.9977 | 0.9987 | 0.9992 | 0.9997 | 1 |
| | 3 | 0.01, 0.13, 0.24, 0.35, 0.41 | 0.7448 | 0.9812 | 0.9657 | 0.9992 | 0.998 | 0.9992 |
| | 4 | 0.01, 0.011, 0.012, 0.02, 0.41 | 0.9896 | 0.9999 | 0.9999 | 0.9999 | 0.9993 | 0.9996 |
| | 5 | 0.01, 0.013, 0.028, 0.23, 0.41 | 0.953 | 0.9999 | 1 | 1 | 1 | 1 |
| Moderate "linear" | 6 | 0.01, 0.01, 0.06, 0.11, 0.16 | 0.9457 | 0.5957 | 0.6499 | 0.7094 | 0.8198 | 0.8729 |
| Moderate Plateau | 7 | 0.01, 0.21, 0.21, 0.21, 0.21 | 0.7417 | 0.2053 | 0.2102 | 0.5478 | 0.4399 | 0.5563 |
| Strong plateau | 8 | 0.01, 0.41, 0.41, 0.41, 0.41 | 0.7409 | 0.3935 | 0.4230 | 0.9997 | 0.891 | 0.9382 |
| Shifted model: Linear, emax, sigEmax, exp, logistic | 9 | 0.26, 0.27, 0.28, 0.33, 0.66 | 0,0004 | 0.9742 | 0.9824 | 0.9775 | 0.8913 | 0.9303 |
| | 10 | 0.26, 0.28, 0.32, 0.43, 0.66 | 0,0003 | 0.9699 | 0.9708 | 0.9737 | 0.9342 | 0.9612 |
| | 11 | 0.26, 0.38, 0.49, 0.6, 0.66 | 0,0001 | 0.9666 | 0.8673 | 0.9699 | 0.9598 | 0.9766 |
| | 12 | 0.26, 0.26, 0.26, 0.27, 0.66 | 0,0005 | 0.9793 | 0.9899 | 0.9833 | 0.8571 | 0.9088 |
| | 13 | 0.26, 0.26, 0.28, 0.48, 0.66 | 0,0004 | 0.9820 | 0.9856 | 0.9854 | 0.9618 | 0.9774 |
| Misspec. model: betaMod, sigEmax, logistic | 14 | 0.01, 0.075, 0.2, 0.4, 0.082 | 0.7858 | 0.8599 | 0.1833 | 0.4281 | 0.6766 | 0.7915 |
| | 15 | 0.01, 0.32, 0.37, 0.39, 0.41 | 0.7409 | 0.6414 | 0.6381 | 0.9996 | 0.9523 | 0.9478 |
| | 16 | 0.01, 0.012, 0.022, 0.27, 0.41 | 0.9616 | 1 | 1 | 1 | 1 | 1 |
| Misspec. Shifted model: betaMod, sigEmax, logistic | 17 | 0.26, 0.32, 0.45, 0.65, 0.33 | 0,0002 | 0.5368 | 0.2011 | 0.5451 | 0.3848 | 0.4790 |
| | 18 | 0.26, 0.57, 0.62, 0.64, 0.66 | 0,0001 | 0.9569 | 0.3976 | 0.9614 | 0.8843 | 0.9261 |
| | 19 | 0.26, 0.26, 0.27, 0.52, 0.66 | 0,0005 | 0.9885 | 0.9895 | 0.9912 | 0.9716 | 0.9838 |